\documentstyle[11pt,psfig]{article}

\begin{document}

\title{\bf \Large A dusty torus around
                  the luminous young star LkH$\alpha$\,101} 

\author{\bf Peter G. Tuthill$^\ast$, John D. Monnier$^\dag$ \& 
            William C. Danchi$^\ddag$}

\date{}

\maketitle

{\small \sl $\ast$
Astronomy Department, School of Physics, Univ of Sydney, Sydney NSW 2006, Australia.}\\
{\small \sl $\dag$
Harvard-Smithsonian Center for Astrophysics, 60 Garden St, Cambridge, MA 02138, USA.}\\
{\small \sl $\ddag$
NASA Goddard Space Flight Center, Infrared Astrophysics, Code 685, 
Greenbelt, MD 20771, USA.}

\bf
A star forms when a cloud of dust and gas collapses.
It is generally believed that this collapse first produces a
flattened rotating disk\cite{Terebey_84,Yorke_95}, through which matter
is fed onto the embryonic star at the center of the disk.
When the temperature and density at the center of the star pass a
critical threshold, thermonuclear fusion begins. The remaining disk,
which can still contain up to $\sim$0.3 times the mass of the 
star\cite{Cassen_81,Shu_90,Hollenbach_94}, is then sculpted and eventually
dissipated by the radiation and wind from the newborn star.
Unfortunately this picture of the structure and evolution of the disk remains
speculative because of the lack of morphological data of sufficient resolution
and uncertainties regarding the underlying physical processes.
Here we present resolved images of a young star, LkH$\alpha$~101  
in which the structure of the inner accretion disk is resolved.
We find that the disk is almost face-on, with a central gap (or cavity)
and a hot inner edge.
The cavity is bigger than previous theoretical predictions\cite{Hillenbrand_92},
and we infer that the position of the inner edge is probably determined by 
sublimation of dust grains by direct stellar radiation, rather than by disk
reprocessing or the viscous heating processes as usually assumed\cite{L-B_74}.
\rm

The Herbig Ae/Be stars, thought to be  pre--main-sequence stars of 
intermediate mass, have generated considerable recent controversy 
over their circumstellar structure.
Accretion-disk models were found to fit the spectral energy distribution
(SED) provided that an optically thin hole or cavity was allowed around 
the central star\cite{Hillenbrand_92,Kessel_98}.
This model drew almost immediate criticism over missing accretion 
luminosity\cite{Hartmann93} and problems with forbidden emission line
profiles not matching expectations\cite{Bohm_Catala_94}.
Further confusion arose from the finding that the SED could
be fitted by spherically-symmetric circumstellar 
shells\cite{Miroshnichenko_97,Pezzuto_97} or composite shell-disk 
models\cite{Miroshnichenko_99}.

LkH$\alpha$~101 is amongst the brightest young stellar objects in the 
near-infrared, despite its location behind considerable line-of-sight 
obscuration (recent estimates of visible extinction $A_v$ range from
9.4\cite{Barsony_90} to 18.5\cite{Hou_97}).
Interferometric observations with the Keck~I telescope capable of imaging
structure on scales of tens of milli-arcseconds\cite{keckmask_00} have 
enabled us to observe the circumstellar environment of LkH$\alpha$~101 in 
the near-infrared at an unprecedented level of detail. 

Images presented in Figure~1, taken in the H and K bands, show the bright
core to be resolved into a circular limb-brightened disk with a marked
asymmetric brightening  along the southwest side.
A region of suppressed flux, or dark hole, is apparent at the center.
LkH$\alpha$~101 is also revealed as a binary star, with a companion at a 
separation of 180\,mas to the E-NE in the 1.65\,$\mu$m map.
This binary component is still visible in the map at 2.27\,$\mu$m, although
it is now much fainter, appearing only at the lowest contour with a 
signal-to-noise ratio around 1.
The relative brightening of the companion from K to H band implies a significantly
hotter apparent spectrum. 

\begin{figure}[ht]
\mbox{
\hspace{-1.0cm}
\psfig{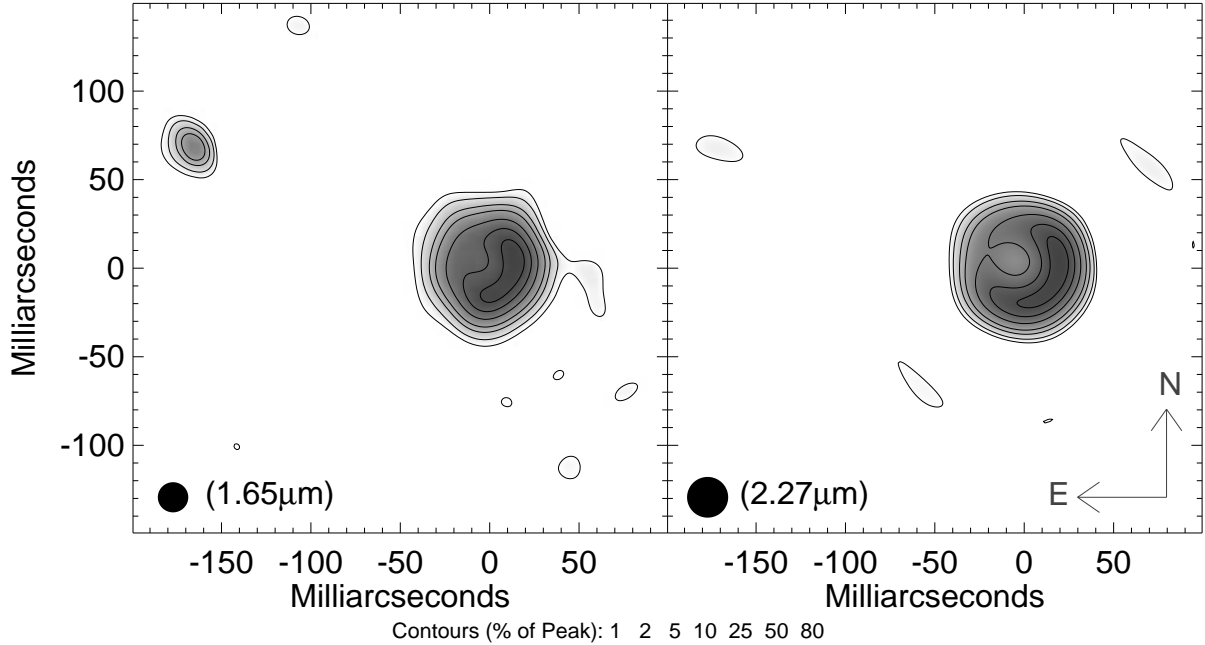}
}
\caption{\sf 
Maps of the LkH$\alpha$~101 system.
Maximum-Entropy image reconstructions at 1.65 (left)
and 2.27\,$\mu$m (right) from data taken in September 1998 (JD~2451086).
Filter bandwidths were 0.33 and 0.16\,$\mu$m respectively. 
As it is difficult to extract both absolute astrometry and photometry with
our interferometric techniques, the origin has been centered on the disk and 
images scaled to the brightest pixel.
The K-band map is of slightly superior quality to H-band due to the 
less-detrimental impact of the seeing at longer wavelengths.
High spatial resolution information was extracted from high magnification 
(20\,mas/pixel) fast-readout ($\sim130$\,ms) data frames taken with
an annulus-shaped aperture mask\cite{keckmask_00}.
Consecutive sets of 100 images, interleaved between LkH$\alpha$~101 and nearby
point-source objects, allowed calibration of the statistical behavior of the
telescope--atmosphere point-spread function.
Data were processed to extract Fourier bispectral data, with the final output
being a calibrated set of 630 Fourier amplitudes and 7140 closure phases.
High-quality images were produced from an algorithm based on the maximum entropy 
method\cite{GS_IEEE,Siv_84}, although other methods such as CLEAN have also
been utilized with near-identical results.
The fidelity of our mapping process has been extensively tested on 
binary stars and other known compact asymmetric targets\cite{wr104,irc+10216}.
}
\label{lkha101}
\end{figure}

The circular appearance of the resolved primary component implies a face-on 
viewing angle onto an accretion disk.
Fitting to the images in Figure~1 gives a maximum possible inclination
angle of about $35^\circ$ to the line-of-sight.
The central star is not seen at either wavelength, as its contribution
is swamped by the extremely high infrared luminosity of the disk.
By examining the visibility curves, we find an upper limit of 5\%
for any unresolved component in the maps.
This should not be surprising as the infrared excesses exhibited by many 
Herbig Ae/Be stars imply that circumstellar material completely dominates 
the SED from the infrared through to the radio wavelengths.

The depression or hole in the emission that we saw at the center of the ring
provides definitive confirmation of the scenario of an accretion
disk with a central optically thin cavity in this young Herbig Ae/Be system.
Although this finding is in accord with the scenario of Hillenbrand et al 
\cite{Hillenbrand_92}, this agreement may be superficial as the cavity 
observed here is much larger than those considered previously.

The asymmetric flux resulting in the bright crescent appearance to the
south-west is a result of the disk being tilted away from the
line-of-sight in this direction.
This allows us to see through the central cavity to the hot inner edge of 
the disk on the far side (in the south-west), while in the north-east our view 
is partially blocked by cooler regions of the disk in the line-of-sight.
Just such a ``peculiar horseshoe-like feature'' was well predicted
\cite{Kessel_98} from models with simulated disks tilted at an 
angle of $30^\circ$, albeit in the mid-infrared ($\lambda=12$\,$\mu$m)
wavelengths.

In order to make meaningful comparison with literature models, we need
to estimate the basic properties of the central star driving LkH$\alpha$~101.
The extremely high absolute bolometric luminosity of 
$4.8\ \times\ 10^4\ L_\odot$\cite{Barsony_90}, where $L_\odot$ is the 
luminosity of the Sun, some 75\% of which is thought be coming from a 
deeply-embedded cluster containing hundreds of objects\cite{Barsony_91}, 
was based on an earlier distance determination of 800\,pc\cite{Herbig_71}.
However, a recent distance estimate mainly based on radio-photometry 
points to 160\,pc\cite{Stine_Oneal_98} -- typical of the Taurus-Auriga 
star formation complex with which LkH$\alpha$~101 may be associated.
If this distance is correct, the implied intrinsic luminosity (a factor of
$\sim$25 times lower ) is $4.8 \times 10^2 L_\odot$ -- more appropriate 
for a mid--late B~star than an O~star as was supposed.
However, radio observations showing an ionized shell\cite{Harris_76,Cohen_82}
imply Lyman continuum photon fluxes such as would result from
a much hotter early B~star (B0 to B1).
Given the uncertainties in the extinction and the known global anisotropies,
an exact classification of LkH$\alpha$~101 is problematic, but assuming
it to be an early B-type zero age main sequence (ZAMS) star gives an approximate
mass $M_\star$ of 5 -- 10\,$M_\odot$, an effective temperature $T_{eff}\sim20\,000$\,K 
and a radius $R_\star \sim4$\,$R_\odot$ \cite{Panagia_73}).

The circular ridge of brightest emission has a radius of 21\,mas measured
from the K-band map of Figure~1.
If we identify this as the hot edge of an inwardly-terminated
accretion disk\cite{Kessel_98}, this implies a physical radius of
3.4\,AU for a distance of 160\,pc (or alternatively 16.8\,AU at 800\,pc).
This is vastly larger than the $3 < R_{hole}/R_\star < 25$ inner cavities 
required by Hillenbrand et al.\cite{Hillenbrand_92} to fit the near-infrared
inflections in the SED.
For example, models of early B~stars from this sample had derived cavity radii 
of 0.3 -- 0.6\,AU, around an order of magnitude smaller than the one we see.
Although we find some qualitative agreement, we are forced to conclude
that our images are inconsistent with classical accretion disk models.

This reinforces findings from recently reported high-resolution
observations of the Herbig objects with the Infrared Optical Telescope
Array (IOTA)\cite{IOTA_ABAur_99,M-G_01} and the Palomar Testbed 
Interferometer\cite{Akeson_00}.
Although most of the visibility data fitted with standard viscously-heated 
accretion disks ($T \propto R^{\frac{-3}{4}}$), an unlikely set of extreme 
inclination angles was required\cite{M-G_01}). 
Instead, simpler spherical Gaussian or ring models worked better, suggesting 
that the ensemble of observations were most consistent with spherical shell 
geometries.
The one-sided banana-shaped brightening on our images does not suggest
spherical symmetry.
The IOTA, with limited sky rotation (typically $\sim 30^\circ$ 
baseline rotation) and Fourier coverage may not have detected similar
asymmetries if present in their targets.
The match between the asymmetric limb-brightened structure we see
and the models of Kessel et al.\cite{Kessel_98} supports models based 
on the idea of an accretion disk with a central cavity for LkH$\alpha$~101.  
Resolved imaging of more systems will be required to settle the conflict
between spherical versus disk models for Herbig Ae/Be stars.

There is a rapidly-developing theoretical framework for interpreting young 
stellar objects at later epochs where the winds and radiation from the 
fully-luminous central star sculpt the remnant accretion disk by 
photo-evaporation and wind ram pressure.
By equating the plasma sound speed with the Keplerian orbital velocity, 
Hollenbach et al.\cite{Hollenbach_94} defined the {\em gravitational radius}
$r_g$ beyond which material is no longer gravitationally bound and
will escape as an ionized wind.
Following these authors, we find for a 7.5\,$M_\odot$ star that  $r_g = 50$\,AU, 
although this does entail extrapolation to less massive stars than were 
originally considered (alternatively $r_g = 100$\,AU if the heavier 
15\,$M_\odot$ stellar mass\cite{Barsony_91} is adopted).
This is significantly larger than the apparent structure we see, implying the
resolved disk fits well within the picture of a gravitationally bound 
circumstellar structure.

The photo-evaporative models \cite{Kessel_98} were constructed
for an 8.4\,$M_\odot$ star surrounded by a 1.6\,$M_\odot$ circumstellar disk --
not dissimilar to our expectations for LkH$\alpha$~101.
Radiative and hydrodynamical modeling resulted in inner disk cavities
around 40\,AU in radius.
Thus the model scale is larger than we observed by almost an order of 
magnitude, although this discrepancy would vanish if the 800\,pc distance 
scale is adopted.
However, the visual match of the appearance of disk models inclined 
at $30^\circ$ \cite{Kessel_98} and the observed limb-brightened crescent of 
Figure~1 encourage hope that the basic structure and physics of radiant
and wind processes describing the sculpting and illumination of the stellar 
disk are on the right track.
Perhaps self-consistent models can be found with closer inner radii.

We proceed with a far simpler comparison of our 3.4\,AU cavity with the radius at
which dust sublimates due to heating from the stellar radiation field:
$R_s = \frac{1}{2}\sqrt Q_R (T_\star/T_s)^2 R_\star$ 
where $T_s$ is the sublimation temperature and $Q_R = Q(T_\star)/Q(T_s)$ the 
ratio of the absorption efficiencies $Q(T)$ of the dust for radiation at 
color temperature $T$ of the incident and reemitted field.
Assuming that grains surviving at the inner edge will be those best able to 
radiate away the energy from the star (blackbody grains) we assume $Q_R = 1$.
Taking $T_s =$1500\,K and using the stellar parameters above gives $R_s = 1.7$\,AU 
-- a reasonable match to our observed cavity given the uncertainties in the 
stellar properties.
Also, our fit was to the peak of the ridge of emission in Figure~1, but
the innermost edge may be somewhat smaller than this.
Finding the dust at (or somewhat beyond) the sublimation radius is
analogous to similar findings for asymptotic giant branch stars, implying that 
the same basic radiative equilibrium processes may set the inner cavity 
size in both cases.

We turn now to the binary companion which was found to contribute 
some 0.8\% of the flux at K and 4.0\% of the flux at H.
Such a luminous (relatively) blue object may be the 
source of Lyman-continuum photoionizing flux.
Alternatively the (unseen) object at the center of the disk may power the
energetic emission-line spectrum.
Given the flux ratios in the near-infrared, it is possible to estimate
the temperature of the companion relative to the disk.
Unfortunately, the uncertainties in the extinction mentioned
above hamper the de-reddening calculations, so we can only 
place limits, finding the companion to be at least 2000\,K hotter
than the disk.

We detected the secondary at multiple epochs and the components 
show no apparent relative motion greater than 15\,mas/yr.
As it seems reasonable to assume that the two objects do form a bound
pair, assuming a total system mass of $\sim 10$\,$M_\odot$ results in
an orbital period of 49\,yr given a physical separation of 28.8\,AU.
The apparent motion exhibited by such a system is comparable to our
limit, and so extra observations at future epochs should reveal the
orbit of this system.
We hope such future observations, and the extension of imaging to 
shorter wavelengths, will help to identify which component provides 
the ionizing flux, powering this young active region.

{\bf Acknowledgements} 
We would like to thank D. Sivia for the maximum-entropy
mapping program ``VLBMEM''.
Data were obtained at the W.M. Keck 
Observatory, made possible by the generous support of the W.M. 
Keck Foundation, operated as a scientific partnership among the
California Institute of Technology, the University of California and NASA.
This work was supported through grants from the National Science Foundation.

Correspondence and requests for materials should be addressed to 
Peter Tuthill \\
(e-mail: gekko@physics.usyd.edu.au)

\end{document}